\documentclass[reqno]{amsart}

\usepackage{amsmath}
\usepackage{amssymb}
\usepackage{amsthm}
\usepackage{amsfonts}
\usepackage{dsfont}
\usepackage{color}
\usepackage{enumerate}
\usepackage[margin=38mm]{geometry}
\usepackage{graphicx}

\usepackage{braket}
\usepackage{hyperref}

\newcommand{\reals}{\mathbb{R}}

\newcommand{\complex}{\mathbb{C}}

\newcommand{\paraa}[1]{\big(#1\big)}
\newcommand{\parab}[1]{\Big(#1\Big)}

\newtheorem{theorem}{Theorem}[section]

\newtheorem{example}[theorem]{Example}
\theoremstyle{definition}

\theoremstyle{remark}

\numberwithin{equation}{section}

\newcommand{\ab}{\bar{a}}

\newcommand{\jb}{\bar{j}}

\title{Noncommutative spaces as quantized constrained Hamiltonian systems}

\author{Andreas Sykora}
\address[Andreas Sykora]{}
\email{syko@gelbes-sofa.de}

\thanks{}

\subjclass[2000]{}

\begin{document}

\begin{abstract}
We investigate the strong-field limit of a charged particle
in an electromagnetic field as a toy model for general covariant systems,
establishing a novel connection between constrained Hamiltonian dynamics
and noncommutative geometry. Starting from the action 
$S=\int d\tau \, \dot{x}^i A_i(x)$, which represents the holonomy
of the particle's path with respect to the electromagnetic potential $A_i$,
we analyze the resulting general covariant system with vanishing Hamiltonian.
The equations of motion $F_{ij}\dot{x}^j=0$ confine the particle to leaves of a singular foliation
defined by the field strength tensor $F_{ij}=\partial_i A_j -\partial_j A_i$.
We show that the physical state space corresponds to the space of leaves of this foliation,
with points connected by field lines being gauge equivalent. The Hamiltonian analysis reveals
constraints $\kappa_i=p_i-A_i$ that are locally classified as first-class or second-class
depending on the rank of the field strength tensor.
Upon quantization, this leads to noncommuting coordinate operators,
establishing the physical state space as a noncommutative geometry.
We provide explicit examples and show in particular that the magnetic monopole field strength
yields a fuzzy sphere. 
\end{abstract}

\maketitle

\tableofcontents

\section{Introduction}

One approach to quantizing gravity is canonical quantum gravity and
in particular loop quantum gravity \cite{Rov04}. General relativity is formulated
as a Hamiltonian system and it turns out that the Hamiltonian of the theory
vanishes and solely constraints remain. Contrary to the quantization of a Yang-Mills
theory, in which time remains as an external parameter, time disappears
at the most fundamental level. The reason for this is the diffeomorphism
invariance of general relativity, which eliminates any preferred notion of time.
This phenomenon is often referred to as "physics without time" and is not
restricted to canonical quantum gravity but occurs in all general covariant
systems, see for example \cite{Hen92}, Chapter 4.

The absence of a global time parameter complicates the interpretation
of the corresponding quantum theory substantially, since the traditional notion
of time evolution is no longer present. 

Noncommutative geometry \cite{Con94} provides a framework for describing “quantum” spaces,
where coordinates do not commute, mirroring the noncommutativity of observables
in quantum mechanics and enabling geometry to survive at Planck-scale regimes,
where classical spacetime notions break down. By replacing functions on spaces with
noncommuting operators, noncommutative geometry extends geometry to settings
that may become relevant for quantum gravity \cite{Ste24, Ste10, FPV23}.

In the present work, we consider the strong field limit of
a charged particle in an electromagnetic field in flat space, and will see
that one can treat the resulting system as a general covariant system.
Additionally it turns out that the physical state space can be interpreted
as a noncommutative geometry.

On the classical side, we will see that the particle is confined to leaves of a foliation
defined by the field strength and the physical state space reduces to a lower dimensional
subspace. In the Hamiltonian theory, the physical state space is provided with a Dirac bracket,
which is non-zero for the configuration space coordinates. Consequently, after quantization,
the physical state space becomes a noncommutative space. This is an interesting
link between general covariant systems and noncommutative geometry.

Our starting point is the following action in flat space $\reals^n$ with arbitrary dimension $n>1$ 
\begin{align}  
  S_{full} =  \int d\tau \, \paraa{ L_{free}(\dot{x}) - q \dot{x}^i A_i(x) }
\end{align}
where $L_{free}$ can be the free Newtonian $\frac{1}{2}m\dot{x}^2$
or relativistic  $m\sqrt{-\dot{x}^2}$ Lagrangian. The particle
has charge $q$ and is minimally coupled to the the potential $A$
of the electromagnetic field. In general, we think of the potential $A$ as the
connection of a $U(1)$ fibre bundle. It is possible to restrict to
subsets of $\reals^n$, which makes it possible to also consider
the potential of a magnetic monopole.

The limit $\frac{m}{q}\rightarrow 0$ of strong electromagnetic fields
results in the action
\begin{align}  \label{lim_action}
  S = \int d\tau \, \dot{x}^i A_i(x) = \int_\gamma A
\end{align}
which is basically the holonomy of the path $\gamma$ of the particle with respect to
the one form $A=A_i dx^i$.
Any metric, which is solely present in $L_{free}$ has dropped out
and the system becomes invariant with respect to coordinate transformations.
Below, we will show that the corresponding Hamiltonian theory is
general covariant and has zero Hamiltonian. In such a way,
it can be considered as a very simple toy model for gravitational theories.

Additionally, the action (\ref{lim_action}) is invariant with respect to world-line reparametrizations
$x(\tau)\mapsto x(\tau(\tau'))$ and with respect
to local gauge transformations $A_i \mapsto a_i+\partial_i \phi$. 
Note that after a gauge transformation, the action for a finite path
adopts $U(1)$-factors at the ends of the path. Therefore, the invariance
with respect to local gauge transformations is only present for infinite or closed paths.

Although the dynamics of the system (\ref{lim_action}) are rather trivial,
it has an interesting physical state space. We will see that the Hamiltonian theory leads to
Dirac brackets that depend on the field strength of the potential $A$, resulting
in noncommutative coordinates after quantization. 

The approach described in the following differs from the usual way in which a particle in
the limit of a strong electromagnet field is quantized. Usually, first the particle in 
the electromagnetic field is quantized and then, for performing the strong 
field limit, a projection to the first Landau level is performed. For example, \cite{Grei11}
mentions the case of a particle in the plane and proposes a similar formalism for Landau level 
quantization on a sphere. Here, we already perform the strong field limit in the 
classical system and quantize afterwards.

More general, in \cite{Dou10} and \cite{Kle09}, the projector on the lowest Landau level,
i.e. the Bergman kernel, is calculated using a path integral for a particle in a strong magnetic field.
This connects to the usual way, how symplectic manifolds compatible with
a complex structure are quantized. In this setting, the quantum Hilbert space
is constructed as the space of holomorphic sections of a positive line bundle
over the Kähler manifold, and the Bergman kernel serves as the reproducing kernel 
for this space of holomorphic functions. In the present approach, we start with
a one form or more general with a connection of a complex line bundle, and quantize
a special covariant Hamiltonian system by finding the constraints and
implement the constraints in the standard Fock space.

The structure of this paper is as follows: In Section 2, we analyze the minimal coupled Lagrangian action
and derive the equations of motion, showing how they relate to singular foliations.
The topology of the physical state space is examined and explicit examples
including the two-dimensional case and magnetic monopole field configurations are provided.

Section 3 develops the Hamiltonian theory. The types of constraints are examined
and the Dirac brackets are derived. It turns out that the notion of
first-class and second-class constraints varies locally.

Section 4 presents a quantization scheme using generalized
Fock space methods. Examples are provided including a "disc",
a "stack of planes" and the case of a monopole field strength. It turns out
that the monopole field strength results in a fuzzy sphere \cite{Mad92, Hop82}.

As a side remark, in \cite{Con94} spaces of leaves of foliations
are provided with a $C^*$-algebra structure. Since in the present
work foliations also arise, the question arises whether the two
approaches have something in common. It turns out that the
two approaches are different. This is discussed in section 5.

\section{Topology of the physical state space}

Varying the action (\ref{lim_action}) with respect to the $x^i$
or evaluating the Euler Lagrange equations of the Lagrangian 
$L=\dot{x}^i A_i(x)$
\begin{align} 
   \frac{\delta S}{\delta x^i} 
    = \frac{\partial L}{\partial x^i} -\frac{d}{d\tau}\frac{\partial L}{\partial \dot{x}^i}
    = (\partial_i A_j) \dot{x}^j  - (\partial_j A_i) \dot{x}^j
\end{align}
results in the equations of motion (EOM) 
\begin{align} \label{EOM}  
   F_{ij}\dot{x}^j=0
\end{align}
where $F_{ij}=\partial_i A_j -\partial_j A_i$ is 
the field strength of $A_i$. These EOM at a first sight
appear rather trivial. When $F_{ij}$ is invertible at a
point $p_0=(x_0^j)$, then the particle is confined to this
point and the single solution of the EOM running through this point is $x^j(\tau)=x_0^j$.
In this case $F_{ij}$ is a symplectic form and
there is a single solution for every point $p_0$ of $\reals^n$.
The space of solutions, i.e. the classical physical state space,
is parametrized by $\reals^n$.

However, as we will see, when $F_{ij}$ is not symplectic, it is
possible that the space of gauge equivalent solutions can have 
much richer topology than the original configuration space
due to further gauge invariances, which relate to the null space of
$F_{ij}$, i.e. the ideal of tangent vectors $v^i$ in the tangent space, which
are annihilated by $F_{ij}$. Every transformation of $F$, which leaves the 
null space invariant, will also not change the equations of motion (\ref{EOM}).
The nature of these gauge invariances will also become clearer
below, when we discuss the first-class constraints of
the corresponding Hamiltonian system.

\subsection{Two-dimensional case} \label{2D_classical}

Before treating the general case, we first consider the
two-dimensional case $n=2$, in which every two-form or field strength
\begin{align*} 
   F = \rho(x,y) dx \wedge dy
\end{align*}
is automatically closed. In regions of $\reals^2$ where $\rho$ is non-zero,
$F$ is invertible and since it is closed, it is also symplectic.

The EOM (\ref{EOM}) reduce to
\begin{align} \label{2D_EOM}
   \rho\dot{x} =\rho \dot{y}=0 
\end{align}
Thus, in regions, where $\rho\neq0$, there is only the solution
of a constant path $x(\tau)=x_0, y(\tau)=y_0$ for every point $(x_0,y_0)$.
We can identify the solutions with the points in these regions.

On the other hand, when $\rho=0$, there are no EOM. The motion of the particle
is unconstrained in such regions. However, when given a physical state at a time $\tau_1$,
the EOM should determine the physical state at every other time $\tau_2$ uniquely. Otherwise,
there is a gauge invariance and two physical states at a time $\tau_2$
are gauge equivalent, when they can be reached by time evolution of
the system from the same physical state at time $\tau_1$. In the present case,
when we take one point $(x_0,y_0)$ inside a connected region defined by $\rho=0$, we can
connect every other point inside this connected region with an arbitrary path with this point.
Since there are no EOM, such a path is a physical solution. It follows that all points within a
connected region where $\rho=0$ are gauge equivalent.

In summary, in regions, where $\rho\neq 0$, there is no gauge freedom,
and in connected regions where $\rho=0$, all point are gauge equivalent.
Thus, in two dimensions, the physical state space is a plane, where connected regions having $\rho=0$
are shrunk to a point. When such regions are simply connected,
there is no topological difference. For multiply connected regions,
spheres are pinched off. For example, when there is one single
annulus-shaped region with $\rho=0$, the resulting space is a plane,
which touches a sphere in one point.

Below we will consider an example, where $\rho=0$ for all points outside
the unit circle. 

\subsection{Three-dimensional case}

In the three-dimensional case $n=3$, the field strength $F$
can be expressed as the magnetic field $\vec{B} =\nabla \times \vec{A}$,
and the EOM reduce to $\dot{\vec{x}} \times \vec{B}=0$.
This means that $\dot{\vec{x}}\parallel \vec{B}$, i.e. the particle
is confined to the magnetic field lines, but its motion along a given field line
is unrestricted. 

As the field strength $F$ has at least rank two, there is a coordinate
system, in which it can be expressed as 
\begin{align} \label {3DF}
   F = \rho(x,y,z) dx \wedge dy
\end{align}
Since we require that $F$ be closed $dF=0$, it follows that $\partial_z \rho=0$,
i.e. $\rho$  in (\ref{3DF}) depends solely on $x$ and $y$.
In this coordinate system, the magnetic field $\vec{B}$ has only one 
component $\rho$ in $z$-direction.

Repeating the argumentation with respect to gauge equivalent points, i.e.
that two points that can be reached from the same original point via solutions
of the EOM are gauge equivalent, we have to identify points, which are on the same
line of the magnetic field or which have the same $z$ coordinate in the special
coordinate system for (\ref{3DF}). Note that this is only possible,
since $F$ does not depend on $z$. 

Additionally, as in the two-dimensional case above,
points where $\rho(x,y)=0$ have to be identified. In such regions, 
there are no magnetic lines, since the magnetic field $\vec{B}$ or
the field strength $F$ is zero. 

Thus, the space of solutions modulo these gauge invariances can be 
parametrized by the lines of the magnetic field. For example, for the field of a magnetic
monopole, this results in a sphere (see below).

\subsection{Singular foliation of the field strength}

In the general case of $n$ dimensions, we see that the EOM (\ref{EOM}) 
do not constrain $\dot{x}$, when it is in the null space of the two-form
$F=F_{ij}dx^i\wedge dx^j=dA$, i.e. the vector space of all vector fields $X$
with $X^i F_{ij}=0$. In the three dimensional case above, these vector fields
are in parallel to the magnetic field lines.

Let us first consider a general $k$-form $\omega$ and restrict later to the case of a two-form.
The vector fields $X$, which form the null space of the $k$-form $\omega$,
i.e. with $i_X\omega=\omega(X,\cdot)=0$ or locally $\omega_{i_1 i_2 \dots i_k}X^{i_1}=0$,
form a distribution $N_\omega \subset T(\reals^n)$ of the tangent space $T(\reals^n)$ of $\reals^n$.
With vector fields $Y_i\in T(\reals^n)$ the exterior derivative of the $k$-form $\omega$ is
\begin{align} 
   d\omega(Y_0,\dots,Y_k) & = \sum_i (-1)^i Y_i\paraa{\omega(Y_0,\dots,\hat{Y}_i,\dots,Y_k)} \\
     & +\sum_{i<j} (-1)^{i+j}\omega\paraa{[Y_i,Y_j],\dots,\hat{Y}_i,\dots,\hat{Y}_j,\dots,Y_k} \nonumber
\end{align}
where $[Y_i, Y_j]$ denotes the Lie bracket
and a hat denotes the omission of the respective vector field.

If the  $k$-form $\omega$ is closed $d\omega=0$, the null space $N_\omega$ is integrable, since then for
$X_0, X_1 \in N_\omega$ and $Y_2,\dots Y_k\in T(\reals^n)$
\begin{align} 
   0 = d\omega(X_0,X_1, Y_2,\dots,Y_k)
   =   - \omega\paraa{[X_0,X_1],Y_2, \dots,Y_k}
\end{align}
i.e. $[X_0,X_1] \in N_\omega$. It follows that the null space $N_\omega$ is a
singular foliation, i.e. a foliation with leaves that can
have different dimensions \cite{Lau22}. In summary, every closed form $\omega$
defines with its null space $N_\omega$ a singular foliation.

Furthermore, the closed form $\omega$ is invariant with respect to the foliation.
When $\omega$ is closed $d\omega =0$  and $X$ is a vector field in the null space $N_\omega$
of $\omega$, i.e. $i_X \omega = 0$, it follows that the Lie derivative vanishes
\begin{align} \label{Lie_der}
  L_X \omega = i_X d\omega + d (i_X \omega) = 0
\end{align}
This means that $\omega$ is constant, when parallel transported along the
flow defined by $X$. Since this is valid for any vector field $X$ in the null space
$N_\omega$, $\omega$ is invariant along the foliation.

When there are local coordinates $(x^i, z^j)$, where the $z^j$ parametrize the
leaves of the foliation, it follows that $\omega$ has the form
\begin{align}  \label{om_loc}
  \omega = \omega_{i_1\dots i_k}(x)  dx^{i_1}\wedge \dots\wedge dx^{i_k}
\end{align}
i.e. $\omega$ does not depend on the $z^j$. This can be shown by using 
the vector fields $X_{z^i}=\partial_{z^i}$ in (\ref{Lie_der}). 

In the case of the closed field strength two-form $F$, the paths, which
are defined by the EOM (\ref{EOM}) are in parallel to the leaves of the null-space foliation.
In the local coordinates of (\ref{om_loc}), the field strength becomes
\begin{align} \label{F_loc}
  F = F_{ij}(x) dx^i \wedge dx^j
\end{align}

$F_{ij}$ needs not be invertible in the region, where the local coordinate
system is defined. There can be singular points
at which the rank of $F$ jumps.
The rank of $F$ at a point is an even number $2p$. At a point, where $F$ 
has constant rank in a region around the point, the leave of the foliation
through this point has dimension $n-2p$, see for example \cite{Lau22}, Theorem 1.6.15.

When we assume that locally in a region $R\subset\reals^n$ the closed 2-form $F$
is of constant rank $2p$, than according to Darboux's Theorem, 
there is a local coordinate system $x^i,i=1,\dots,p$, and $y^i,i=1,\dots,p$
optionally with further $z^i,i=1\dots,N-2p$, such that 
\begin{align} \label{F_spec_sol}
  F= \sum_{i=1}^{p} dx^i \wedge dy^i
\end{align} 
In this case, a one-form with $F=dA$ is $A= \sum_{i=1}^{p} x^i \, dy^i$.

\begin{figure}[h]
  \centering
  \includegraphics[height=5cm]{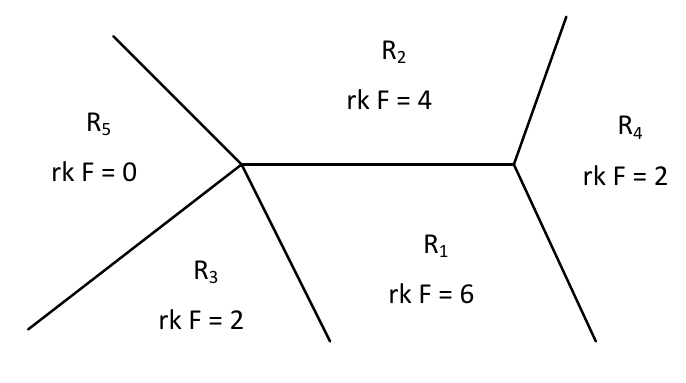}
  \caption{A schematic drawing of regions of different rank of the field strength $F$}
  \label{fig:regions}
\end{figure}

\subsection{The physical state space}

As already explained, when $\dot{x}^i$ is in the nullspace of the field strength $F$,
the EOM (\ref{EOM}) do not constrain $\dot{x}^i$. Otherwise, the 
EOM (\ref{EOM}) demand that $\dot{x}^i=0$. The nullspace of $F$ defines
a foliation and since all pairs of points of one leave of the foliation can be connected
by a solution of the (\ref{EOM}), every leave of the foliation is only one point
in the physical state space. Thus, the physical state space is the space of
leaves of the singular foliation defined by the nullspace of the field strength $F$.

Locally, the physical state space can be parametrized by the $x^i$ in (\ref{F_loc})
or the $x^i$ and $y^i$ in (\ref{F_spec_sol}). 

Considering the complete space $\reals^n$ as original configuration space,
the two-form $F$ can have varying rank $2k$ jumping between the even
numbers $0,2,\dots,2p$, where $2p\leq n$ is the maximum of the rank of $F$.
In other words, $F$ defines a function $\text{rk}\,F:\reals^n\rightarrow\{0,2,\dots,2p\}$,
which sections $\reals^n$ into regions $R_i$ ($i\in I$ an index set),
where the function $\text{rk}\,F$ is constant.

As we have seen above, in every such region $R_i$, the constant rank
$\text{rk}\,F=2k$ of $F$ is the dimension $2k$ of the set of physical states
$R_{\text{phys},i}$.  By applying the equivalence relation
of gauge equivalent points, regions $R_i$ with $\text{rk}\,F=0$ are shrunk
to a point $R_{\text{phys},i}$,
regions $R_i$ with $\text{rk}\,F=2$ are shrunk to a subset of a 
two-dimensional surface $R_{\text{phys},i}$, and in general regions
$R_i$  with $\text{rk}\,F=2k$ are shrunk to a subset $R_{\text{phys},i}$ of
a $2k$-dimensional submanifold. What remains is a collection of topological polyeders.
The regions $R_i$ of maximal rank $\text{rk}\,F=2p$ are shrunk to volumes $R_{\text{phys},i}$
of dimension $2p$, which are bordered by regions $R_i$ 
of lower rank, which result in lower dimensional volumes $R_{\text{phys},i}$ 
of dimension $2k$ with $k<p$. 

Fig. 1 schematically shows, how the configuration space is shrunk to the region $R_1$, in which
the field strength has maximal rank $2p=6$ and which becomes a 6-dimensional part
$R_{\text{phys},1}$ of the physical state space. The other regions $R_i$ become
borders of $R_{\text{phys},1}$. For example, $R_2$ becomes a 4-dimensional border
$R_{\text{phys},2}$ of $R_{\text{phys},1}$.

It has to be remarked that when we start from a punctured $\reals^n$ as configuration space,
such as in the case of a monopole field for which the origion is excluded,
it is possible that the shrunk regions $R_{\text{phys},i}$ alone can have non-trivial topology, such as a sphere.

In the following we work out three examples, which we will also consider
in the Hamiltonian theory and will quantize in the end.

\begin{example}Pinched off disc \label{POD} \end{example} 

This example illustrates the case, where the rank of $F$ jumps
such that a leave of the foliation has a border.

We parametrize the plane in polar coordinates $(r,\varphi)$
and define a field strength by $F = d\rho\wedge d\varphi=\rho'(r) dr \wedge d\varphi$ for $r\leq r_0$ and $F=0$ for $r>r_0$.
$\rho'$ is the $r$-derivative of a function $\rho$ in $r$, wherein $\rho'$ is non-zero inside the disc.
For example $\rho=r^2/2$ for the standard flat symplectic form $dx\wedge dy$ in polar
coordinates. In summary, the field strength is non-zero within the disc of radius $r_0$ and $0$ outside. 

A possible potential is $A= \rho(r) \, d\varphi$ for $r\leq r_0$ and $A=0$ for $r>r_0$.

The singular foliation has one two-dimensional leave $r>r_0$, where the rank of
$F$ is $0$. All points inside the disc are zero-dimensional leaves, where the
rank of $F$ is $2$.

Within the disc $r<r_0$, every point $(r,\varphi)$ corresponds to a solution
of the EOM (\ref{2D_EOM}). Outside of the disc $r>r_0$, all points
have to be identified due to gauge equivalance.
The classical physical state space is a topological sphere.

\begin{example}Stack of planes \label{SOP} \end{example} 

This exemplifies the case, where the rank of the field strength is
locally constant within a neighborhood of a point. In such a region,
we always can define local coordinates where $F$ is constant, see (\ref{F_spec_sol}).

We consider $n=3$ and $\text{rk}\,F = 2$, i.e. $\reals^3$ with coordinates $x,y,z$,
with constant field strength $F= dx\wedge dy$.
A possible potential is $A=-\frac{y}{2}\,dx+\frac{x}{2}\,dy$.

The null space is in parallel to the coordinate $z$.
The singular foliation is composed of the lines parametrized by $(x=x_0, y=y_0, z)$
with $x_0$ and $y_0$ constant and $z$ in $\reals$. Along these lines,
the field strength $F$ is constant. All points on a line are gauge
equivalent. The classical physical state space is the space of these lines
and can be identified with $\reals^2$ parametrized by $(x_0, y_0)$.
$F$ restricted to the physical state space is the constant symplectic form on the plane.

\begin{example}Magnetic monopole field strength \label{MMFS} \end{example} 

To provide an example with nontrivial topology, we consider 
in $\reals^3\setminus \{0\}$ the magnetic monopole field strength with 
integer charge $N$
\begin{align} 
  F = \frac{N}{2}\epsilon^{ijk}\frac{x^i}{r^3}dx^j\wedge dx^k = \frac{N}{2} \sin \vartheta d\vartheta \wedge d\varphi
\end{align}
see \cite{Shn05}.
(The line in $\reals^3$ defined by $\sin \vartheta=0$ is a coordinate singularity.)

$F$ is the field strength of a line bundle with the local Dirac potentials 
\begin{align} \label{Dir_pol}
  A_\pm = \frac{N}{2}\frac{1}{r} \frac{1}{z\pm r} (x dy-y dx) = \frac{N}{2}(\pm1-\cos\vartheta)d\varphi
\end{align}
$A_+$ and $A_-$ are the potentials on the coordinate patches 
$\reals^3\setminus \{z+r=0\}$ and $\reals^3\setminus \{z-r=0\}$.
On the overlap of these two coordinate patches, the two potentials are related
by the infinitesimal gauge transformation $A_+-A_-=Nd\varphi$ corresponding
to the group-valued gauge transformation $e^{iN\varphi}$, which is only continuous,
when $N$ is an integer. 

In spherical symmetric coordinates the null space is in parallel to the coordinate $r$. 
The magnetic lines are rays starting at the origin. The magnetic field $\vec{B}$ depends only on
$\vartheta$ and in particular not on $r$. This results in a singular foliation
of $\reals^3\setminus \{0\}$ composed of all rays starting at the origin.
There is gauge invariance in the $r$-coordinate. The classical physical state space 
can be parametrized by the points of a single sphere.

The EOM $F_{ij}\dot{x}^i =0$ result in $\dot\varphi=0$ and $\dot\vartheta=0$,
which also shows that the classical physical state space is a sphere parametrized
by $\varphi$ and $\vartheta$. $F$ restricted to the physical state space is
the constant symplectic form on the sphere.

\section{Hamiltonian theory}

Varying the Lagrangian (\ref{lim_action}) with respect to $\dot{x}^i$ results in
the canonical momenta $p_i=A_i$ and in the primary constraints
\begin{align}  \label{constr}
   \kappa_i = p_i - A_i =0
\end{align}
The Hamiltonian
\begin{align}  
   H = p_i \dot{x}^i - L = \dot{x}^i  (p_i  - A_i ) = 0
\end{align}
 is zero, which also shows that (\ref{lim_action})
 is a general covariant system. Since $\{H,\kappa_i\}=0$ trivially,
 there are no secondary constraints. $\{\cdot,\cdot\}$ denotes
 the canonical Poisson bracket with $\{x^i,p_j\}=\delta^i_j$.

The Hamiltonian EOM can be derived from the extended action
\begin{align}  \label{Ham_EOM}
  S_H = \int d\tau \, (p_i \dot{x}^i - v^i\kappa_i) = \int d\tau \,  \paraa{ p_i  (\dot{x}^i - v^i) - v^i  A_i }
\end{align}
with Lagrange mutlipliers $v^i$. One can verify that
the Lagrangian EOM (\ref{EOM}) are derivable from
the Hamiltonian EOM (\ref{Ham_EOM}).

\subsection{First-class and second-class constraints}

In general, one further distinguishes first-class constraints
and second-class constraints. A constraint is first-class, when
its Poisson bracket with all other constraints vanishes weakly, i.e.
is a linear combination of the constraints. Constraints without
this property are second-class. For the constraints (\ref{constr})
the Poisson bracket is
\begin{align} \label{gen_com}
   \{\kappa_i,\kappa_j\} = - \partial_{x^i} A_j + \partial_{x^j} A_i = - F_{ij}
\end{align}
In the analysis above, we have seen that we can find a local coordinate system,
in which the field strength form (\ref{F_loc}) solely depends on $2p = \text{rk}\,F$ coordinates
$x^i$ and that the two-form $F$ is constant with respect to $n-2p$ further coordinates $z^i$.
In this coordinate system (\ref{gen_com}) becomes 
\begin{align}  \label{spec_com}
   \{\kappa_{x^i}, \kappa_{x^j}\} = - F_{ij}, \qquad
   \{\kappa_{x^i}, \kappa_{z^k}\} = \{\kappa_{z^k}, \kappa_{z^l}\} = 0 
\end{align}
for $i,j=1,\dots,2p$ and $k,l=1,\dots,n-2p-1$,
where $\kappa_{x^i}=p_{x^i}-A_{x^i}$ and $\kappa_{z^k}=p_{z^k}$
are the constraints related to the coordinates $x^i$ and $z^k$, respectively.

Here and in the following, we will use a notation, where we index quantities
with indices that are the corresponding coordinates. This mean that for
example $\kappa_{x^i}$ means the constraint, which is associated with the
same index as the coordinate $x^i$. $p_{x^i}$ does not depend on $x^i$ but
has the same index as $x^i$. This simplifies to distinguish between constraints,
momenta, partial derivates, etc. which are associated with coordinates, that
are grouped, such as $x^i$, $y^i$, $z^j$, for example, where the index $i$ is
from a different index set as the index $j$.

In a region where the rank of $F$ is constant, see (\ref{F_spec_sol}),
we can use Darboux coordinates $x^i$, $y^i$, $z^j$ with $i=1,\dots,p$ and $j=1,\dots,n-2p$,
and (\ref{spec_com}) reduces to $\{\kappa_{x^i},\kappa_{y^j}\} = \delta^{ij}$,
while all other brackets vanish. 

From (\ref{spec_com}) follows that the $\kappa_{x^i}$
are $2p$ second-class constraints, when $F_{ij}$ is non-zero, and the $\kappa_{z^k}$ are
$n-2p$ first-class constraints. It is not possible to globally asign a
constraint (\ref{constr}) to be first-class or second-class. In general,
it is only possible to state that at a point, which is in a region where the two-form $F$ has
constant rank $2p$, the constraints (\ref{constr}) contain locally $2p$ second-class
and $n-2p$ first-class constraints.

Above we have seen that the leaves of the foliation, which are locally parametrized
by the coordinates $z^k$, correspond to gauge invariant points.
This is confirmed by (\ref{spec_com}), since it is well known that
first-class constraints are generators of gauge-transformations. However,
in the present cause, there is no gauge Lie algebra but solely a gauge
Lie algebroid defined by the vector fields of the null space $N_F$ of $F$.

\subsection{Dirac bracket}

To treat system with second-class constraints, Dirac introduced
the Dirac bracket, which is compatible with the first-class
constraints.

In the local coordinate system of (\ref{spec_com}) in a region where the rank
of $F$ is constant, the matrix $F_{ij}$ is invertible, since it has maximal rank $2p$.
The Dirac bracket there becomes
\begin{align} \label{di_bra}
  \{f,g\}_{DB} = \{f,g\} + \{f,\kappa_{x^i}\} \theta^{ij} \{\kappa_{x^j},g\}
\end{align}
where $\theta^{ij}$ is the inverse matrix of the matrix $F_{ij}$. $f$ and $g$ are two functions
on phase space and in general depend on all coordinates $x^i$ and $z^k$
and their momenta $p_{x^i}$ and $p_{z^k}$. (The unusual $+$ in 
(\ref{di_bra}) is due to the $-F_{ij}$ in (\ref{gen_com}). )

At points, where the rank of $F$ changes, it is not possible to define the
Dirac bracket. There, components of $F$ become zero and $\theta^{ij}$
necessarily diverges. This contrasts with a Poisson manifold,
where $\theta^{ij}$ is defined globally and can become zero. 

Since the Dirac bracket (\ref{di_bra}) vanishes on any constraint
$\{\kappa_{x^i},f\}_{DB}=\{\kappa_{z^k},f\}_{DB}$ for any function
$f$ on phase space, it is possible to restrict it to the physical state space
and to consider solely functions on the physical state space, which do
not depend on the coordinates $z^k$ and their momenta $p_{z^k}$, i.e.
to gauge invariant functions.

We are then able to compute the Dirac bracket on the physical phase
space locally parametrized by the $x^i$ and their momenta $p_i = p_{x^i}$.
Since the constraints commute with all functions $f$, i.e. $\{\kappa_i ,f\}_{DB}=0$, it follows
that $\{p_i,f\}_{DB}=\{A_i,f\}_{DB}$. Thus
\begin{align} \label{DB}
  \{x^i,x^j\}_{DB}  & = \theta^{ij} \\ \nonumber
  \{x^i,p_j\}_{DB}  & =  \{x^i, A_j\}_{DB} = \theta^{ik} \partial_k A_j = \delta^i_j + \theta^{ik}\partial_j A_k \\
  \{p_i,p_j\}_{DB}  & =  \{A_i, A_j\}_{DB}  = \theta^{kl} \partial_k A_i  \partial_l A_j =\theta^{kl} \partial_i A_k  \partial_j A_l \nonumber
\end{align}
where the last step in the second and third line
follows from $\theta^{ik}F_{kj}=\delta^i_j$.

In a Darboux coordinate system, where $F_{ij}$ and $\theta^{ij}$
are constant and $A_j=\frac{1}{2}F_{ij}x^i$ is a solution for the gauge
potential, the constraints are $p_i-\frac{1}{2}F_{ij}x^i$ and the relations reduce to
\begin{align}
  \{x^i,x^j\}_{DB}  = \theta^{ij}, \qquad
  \{x^i,p_j\}_{DB}  =  \frac{1}{2} \delta^i_j, \qquad
  \{p_i,p_j\}_{DB}  = -\frac{1}{4} F_{ij} 
\end{align}

When we are able to quantize the system (\ref{lim_action}), we know that
in the semi-classical limit the commutators will become the Dirac bracket.
This shows that in the quantized system, the coordinate functions will have
a commutator, which up to first order is the (pseudo-)inverse of the field strength $F$.

\begin{example}Pinched off disc\end{example}
Continuing example \ref{POD}, the two constraints are
\begin{align}
  \kappa_r = p_r, \qquad \kappa_\varphi = p_\varphi - \rho(r)
\end{align}
The Poisson bracket of the two constraints is
\begin{align}
  \{ \kappa_r, \kappa_\varphi \} = \rho'
\end{align}
which is non-zero inside the disc $r<r_0$ and $0$ outside. Therefore,
the constraints are second-class inside the disc and first-class outside.
Outside the disc, there is a gauge symmetry.

The Dirac bracket for the physical states parametrized by the points
inside the disc is
\begin{align}
  \{ r, \varphi \}_{DB} = \frac{1}{\rho'}
\end{align}
i.e. diverges on the border of the disc.

\begin{example}Stack of planes\end{example}

In the example \ref{SOP}, the constraints are
\begin{align}
  \kappa_x = p_x + \frac{y}{2}, \qquad \kappa_y = p_y - \frac{x}{2}, \qquad \kappa_z = p_z
\end{align}
$\kappa_x$ and $\kappa_y$ are second-class constraints, while $\kappa_z$ 
is a first-class constraint and there is a gauge symmetry in $z$ direction.

The Dirac bracket for the physical states parametrized by the points of the
$x,y$-plane is
\begin{align}
  \{ x, y  \}_{DB} = 1
\end{align}

\begin{example}Magnetic monopole field strength\end{example}

For example \ref{MMFS}, the constraints in spherical symmetric
coordinates are
\begin{align} 
  \kappa_\varphi = p_\varphi + \frac{N}{2}(\cos\vartheta\pm1), \qquad \kappa_\vartheta = p_\vartheta, \qquad \kappa_r = p_r
\end{align}
$\kappa_\varphi$ and $\kappa_\vartheta$ are second-class, while $\kappa_r$ is
first-class. There is a gauge symmetry in $r$ direction, reducing the physical state space
to a sphere parametrized by $\varphi$ and $\vartheta$.
The Dirac bracket for the physical states parametrized in these coordinates is
\begin{align}
  \{ \varphi, \vartheta  \}_{DB} = \frac{2 }{N \sin\vartheta}
\end{align}
This diverges at the poles of the sphere, which however is related to a coordinate singularity.

\section{Quantization}

We have shown that the field strength $F$ is invertible on the physical
state space. When the physical state space is a manifold,
the field strength transforms it into a symplectic manifold. 
There are several known methods, how to quantize symplectic
manifolds, such as for example geometric quantization \cite{Bat92}, \cite{Sch21}
and Berezin-Toeplitz quantization \cite{Bor94}, \cite{Ma08}, \cite{Sch10}.

Here, we persue a different approach. We start with a Hilbert space of functions
(or more general the sections of a line bundle), in which the 
operators $\hat{p}_i$ and $\hat{x}^i$ are realized as canonical pairs
and try to restrict to a smaller Hilbert space, the quantum physical state space,
in which the constraints are implemented, i.e. $\kappa_i \phi = 0$ for a
states $\phi$. 

One sees immediately that for the ordinary Schrödinger
representation this results in a one-dimensional quantum physical state space, since
the $n$ constraints $\-i\hbar\partial_i + A_i$ applied to a space
of functions in $\reals^n$ reduce the degrees of freedom
to $0$.

However, when we are able to find a coordinate system,
in which the $A_i$ form the connection of a Kähler potential, then this
approach has non-trivial solutions, as we will show in the following.

\subsection{Generalized Fock space quantization}

In particular, we will apply a kind of generalized Fock space quantization.
(see for example \cite{Hen92}, Chapter 13.4).

We start with the obersvation that the number of constraints $\hat{\kappa}_{x^i}$ in (\ref{spec_com})
is even. Therefore, it may be possible to find
linear combinations of these constraints,
resulting in in pairs of constraints, $\hat{\kappa}_i, \hat{\kappa}'_i$ with $i=1,\dots, p$,
which are Hermitian anti-conjugate to each other $\hat{\kappa}'_i=-\hat{\kappa}_i^\dagger$.
Mathematically this means that the manifold has a complex structure.

With this, it turns out that for each pair of constraints, it is possible to consider solely one of the 
constraint $\hat{\kappa}_i=0$ during quantization. In particular, for matrix elements of physical states
with $\hat{\kappa}_i \psi'=0$ it follows that 
\begin{align}  
   <\psi', \hat{\kappa}_i^\dagger \psi> = - <\hat{\kappa}_i \psi', \psi> = 0
\end{align}
This means that in the subspace defined by $\hat{\kappa}_i=0$, $i=1,\dots, p$
also the conjugate constraints are fulfilled. 

To take advantage of this, we assume that the
field strength $F$ is based on a Kähler potential $\phi(a^i,\ab^i)$,
which depends on the complex coordinates $a^i=x^i+i y^i$ and their conjugate complex
$\ab^i=x^i-i y^i$.
$x^i$ and $y^i$ (for $i=1,\dots,p$ where $2p\leq n$) are pairs of real coordinates,
which are combined into the the complex coordinates $a^i$. Note that by an abuse
of notation the $i$ after the plus sign is the imaginary unit, while the $i$
indexing the coordinates is a natural number. Locally, the field strength is
\begin{align} 
  F =  \frac{i}{2}\partial_{a^i} \partial_{\ab^j} \phi \, da^i \wedge d\ab^j
\end{align}

A possible one-form $A$ with $F=dA$ is then
\begin{align} 
  A = \frac{i}{4}\paraa{\partial_{\ab^i}\phi \, d\ab^i - \partial_{a^i}\phi \, da^i }
\end{align}
The classical constraints (\ref{constr}) then become
\begin{align} \label{q_constr}
  \kappa_{a^i} = p_{a^i} +  i \partial_{a^i}\phi , \qquad 
  \kappa_{\ab^i} = p_{\ab^i} -  i \partial_{\ab^i}\phi, \qquad 
  \kappa_{z^j} = p_{z^j} 
\end{align}
for $i=1,\dots,p$, $j = 1,\dots,n-2p-1$ where
$p_{a^i}=\frac{1}{2}(p_{x^i} - ip_{y^i})$ and $p_{\ab^i}=\frac{1}{2}(p_{x^i}+ip_{y^i})$.
Remember that the $z^i$ are the coordinates parametrizing the null foliation, see (\ref{F_loc}).
Importantly, the two constraints $\kappa_{a^i}$ and $\kappa_{\ab^i}$
are conjugate complex.

For quantization, we use the Hilbert space of square integrable functions in $a^i$
and $\ab^i$ with the standard inner product. 
\begin{align} \label{sip}
  <\psi',\psi> = \int d^{2p}a \, \overline{\psi'(a,\ab)}\psi(a,\ab)
\end{align}
where $\psi'$ and $\psi$ are two functions in $a$ and $\ab$.
In this space, 
$\hat{p}_{a^i} = -i\hbar \partial_{a^i}$
resulting in the quantized constraints 
\begin{align} 
  \hat{\kappa}_{a^i} = \hbar \partial_{a^i} - \partial_{a^i} \phi, \qquad
  \hat{\kappa}_{\ab^i} = \hbar \partial_{\ab^i} + \partial_{\ab^i} \phi, \qquad
  \hat{\kappa}_{z^j} = \partial_{z^j}
\end{align}
which for $i=1,\dots,p$ are anti-conjugate with respect to each other
$(\hat{\kappa}_{a^i})^\dagger=-\hat{\kappa}_{\ab^i}$,
since $(\partial_{a^i})^\dagger=-\partial_{\ab^i}$
for the standard inner product. It should be emphasized
that here the use of the standard inner product (\ref{sip}) with constant weight
is the only choice, because otherwise $(\partial_{a^i})^\dagger$ would not
be the anti-conjugate of $\partial_{\ab^i}$.
Additionally, (\ref{sip}) is compatible with the quantized constraint $\hat{\kappa}_{z^j}\psi=0$.

States with $\hat{\kappa}_{\ab^i}\psi=0$ are
\begin{align} 
  \psi(a,\ab) = e^{-\frac{1}{\hbar}\phi(a,\ab)} \tilde{\psi}(a)
\end{align}
Restricted to these functions (or more general sections of a line bundle with connection one form $A$)
the Hilbert space inner product (\ref{sip}) becomes
\begin{align}  \label{inner_product}
  <\psi',\psi> = \int d^{2p}a \, e^{-\frac{2}{\hbar}\phi(a,\ab)} \overline{\tilde{\psi}'(a)}\tilde{\psi}(a)
\end{align}
and the restricted Hilbert space can be identified with the Hilbert space 
of holomorphic sections $\tilde{\psi}(a)$ with an inner product,
which has weight $e^{-\frac{2}{\hbar}\phi(a,\ab)}$.

\subsection{Classical limit}

To determine the classical limit of the Hilbert space defined by (\ref{inner_product}),
one can define Toeplitz operators $T_f$ by projecting multiplication by a function $f$
onto this space, and then extracting the star product $f\star g$ as an asymptotic expansion
of the operator product $T_f T_g$ in inverse powers of $\hbar$. 

In \cite{Syk25}, it was shown
that a $\star$-product constructed in this way results in a Poisson bracket
\begin{align} \label{cl_lim}
  \{f, g\} = \theta^{i\jb} \partial_{a^i} \partial_{\ab^i} + \mathcal{O}(\hbar)
\end{align}
where $\theta^{i\jb}$ is the inverse of the Kähler metric $F_{i\jb}=\partial_{a^i} \partial_{\ab^j}\phi$.
In particular in the present case and in the notation of \cite{Syk25} we can set $\mu= 1/g$ in formula (1.6)
of \cite{Syk25}, where $g=\det F_{i\jb}$. (\ref{cl_lim}) then follows from formula (1.16) in \cite{Syk25}.

This shows that up to first order the $\star$-product commutator is 
equal to the Dirac bracket (\ref{DB}).

\subsection{Radial symmetric Kähler potential}

To derive explizit formulas for our examples,
we assume in the following that $\phi(a,\ab)=\phi(a\ab)$ is radially
symmetric. In this case it is possible to
determine the inner product of monomials in $a$ explicitly
\begin{align}  
  <a^n, a^m>  & =  \int d^2 a \, e^{-\frac{2}{\hbar}\phi(a\ab)} \, \ab^n a^m
    = \int  r \, d\varphi dr \,  e^{-\frac{2}{\hbar}\phi(r^2)} r^{n+m} e^{i\varphi(m-n)} \\
      & = \pi \delta^{nm} \int_0^\infty dx \,  e^{-\frac{2}{\hbar}\phi(x)} x^n
        = \pi \delta^{nm} c_n   \label{cn_def}
\end{align}
with $x=a\ab=r^2$, where we have introduced positive real constants $c_n$, which solely
depend on $\hbar$ and $\phi$. (We consider only those $n$, for which the integral converges.)
Thus, 
\begin{align}  
  \varphi_n(a) = \frac{1}{\sqrt{\pi c_n}} a^n
\end{align}
is an orthonormal basis for the Hilbert space.

The matrix elements of an operator $Q(f)$, which is the multipication by the function $f$, are
\begin{align}  \label{mat_el}
  Q(f)_{nm}  = <\varphi_n, f \varphi_m> 
    & = \frac{1}{\pi \sqrt{c_m c_n}} \int d^2 a \, e^{-\frac{2}{\hbar}\phi(a\ab)} \, \ab^n f(a,\ab) a^m
\end{align}

(In general, when $\varphi_n$ is a basis of $H$, then $A\varphi_n=\varphi_m A_{mn}$ for a matrix $A$.
It follows that $AB\varphi_n=\varphi_m A_{mp}B_{pn}$. The matrix coefficents can be
determined by $<\varphi_m, A\varphi_n>
 = A_{mn}$.)
 
The multiplikation with $a$ becomes a raising operator $\hat{a}=Q(a)$
\begin{align}  \label{raising_op}
  \hat{a} \varphi_n(a) = a \varphi_n(a) 
    = \sqrt{\frac{c_{n+1}}{ c_n}}  \frac{1}{\sqrt{\pi c_{n+1}}} a^{n+1}
    = \sqrt{\frac{c_{n+1}}{ c_n}} \varphi_{n+1}(a) 
\end{align}
Since the multiplikation with $\ab$ is the Hermitian conjugate of multiplikation with
$a$, see (\ref{mat_el}), $\hat{\ab}=Q(\ab)=\hat{a}^\dagger$ is a lowering operator
\begin{align} \label{lowering_op}
  \hat{\ab} \varphi_n = \sqrt{\frac{c_{n}}{ c_{n-1}}} \varphi_{n-1} 
\end{align}
It follows that the commutator is diagonal
\begin{align}  \label{commutator}
  [\hat{a}, \hat{\ab}] \varphi_n = \parab{\frac{c_{n}}{ c_{n-1}}-\frac{c_{n+1}}{ c_{n}}} \varphi_{n} 
\end{align}

\begin{example}Pinched off disc\end{example}

Continuing example \ref{POD} of the pinched off disc $F=0$ for $r>r_0$,
we restrict the inner product  (\ref{inner_product}) to the disc
of radius $r_0$.
\begin{align} 
  <\psi',\psi> = \int_0^{r_0} r dr \, e^{-\frac{1}{\hbar}r^2} \int_0^{2\pi} d\varphi \,  \tilde{\psi}'(re^{-i\varphi}) \tilde{\psi}(re^{i\varphi})
\end{align}
We have additionally choosen $\phi=\frac{1}{2}r^2= \frac{1}{2}a\ab$, which is a solution
for the constant field strength $F= \frac{i}{2}da\wedge d\ab = r dr\wedge d\varphi$.

The constants $c_n$ (\ref{cn_def}) become the lower incomplete gamma function
\begin{align} 
  c_n = \int_0^{\sqrt{r_0}} dx \, e^{-\frac{1}{\hbar}x} x^n
       =  \frac{n!}{\hbar^n} \parab{1-e^{-\frac{1}{\hbar}\sqrt{r_0}}\sum_{k=0}^n \frac{ \sqrt{r_0}^k }{\hbar^k k!} }
\end{align}
In the limit of small $\hbar$, the commutator becomes
\begin{align} 
   [\hat{a}, \hat{\ab}] = - \hbar + \mathcal{O}\parab{ e^{-\frac{1}{\hbar} r^2} }
\end{align}
For a correct quantization, one would expect that the commutator
is exactly $-\hbar$ and that the number of basis states is finite,
since a sphere (which should be the classical limit) has finite area.

\begin{example}Stack of planes\end{example}

For the stack of planes of example \ref{SOP}, we 
consider in $\reals^3$ with coordinates $x,y,z$ the
potential $A=-\frac{y}{2}\,dx+\frac{x}{2}\,dy$, which
has constant field strength $F= dx\wedge dy$.
The quantized constraints (\ref{q_constr}) are
\begin{align}  
  \hat\kappa_x = \partial_x +\frac{iy}{2\hbar}, \qquad 
  \hat\kappa_y = \partial_y - \frac{ix}{2\hbar}, \qquad 
  \hat\kappa_z = \partial_z, \qquad 
\end{align}
Using complex coordinates $a=x+iy$ ($\partial_a=\frac{1}{2}(\partial_x-i\partial_y)$ results in 
\begin{align}  
  \hat\kappa_a = \partial_a -\frac{\ab}{4\hbar}, \qquad 
  \hat\kappa_{\ab} = \partial_{\ab} + \frac{a}{4\hbar}, \qquad 
  \hat\kappa_z = \partial_z, \qquad 
\end{align}

For the quantizations scheme (\ref{sip}), the constraint $\hat\kappa_z$
automatically becomes $0$. Due to the gauge invariance with respect to $z$,
all the planes orthogonal to $z$ are identfied.

The other two constraints result in ordinary Fock space quantization.
With $r_0 \rightarrow \infty$ in the previous example,
we deduce that
\begin{align} 
  c_n =  \frac{n!}{\hbar^n} 
\end{align}
and
\begin{align} 
   [\hat{a}, \hat{\ab}] = - \hbar
\end{align}

\begin{example}Magnetic monopole field strength\end{example}

We continue the example \ref{MMFS} with the monopole field strength $F$ of charge $N$.
To describe the respective Kähler structures of the nested spheres,
we first describe two stereographic projections of the nested spheres to a stack of complex planes.

We define coordinate transformations from $\reals^3 \setminus \{r \pm z \neq 0\}$
to $\complex \times \reals^+$ by
\begin{align} \label{raa_coord}
  r = \sqrt{x^2+y^2+z^2}, \qquad a_\pm = \frac{x+iy}{r \pm z}, \qquad  \ab_\pm = \frac{x-iy}{r \pm z}
\end{align}
These coordinate transformations are not defined for $r\pm z =0$ and can be considered as 
a mapping for the coordinate patches of the line bundle for the monopole field such as described in
example \ref{MMFS}.

From (\ref{raa_coord}) follows
\begin{align} 
  a_\pm\ab_\pm= \frac{x^2+y^2}{(r \pm z)^2}, \qquad 1+a_\pm\ab_\pm = \frac{2r}{r \pm z}
\end{align}
and therefore the back transformation is
\begin{align} \label{back_trans}
  x = r\frac{a_\pm+\ab_\pm}{1+a_\pm\ab_\pm}, \qquad 
  y = -i r\frac{a_\pm-\ab_\pm}{1+a_\pm\ab_\pm}, \qquad 
  z = \mp r \frac{a_\pm\ab_\pm-1}{1+a_\pm\ab_\pm}
\end{align}
From this follows that the same point $(x,y,z)$ is mapped to
the points $(r,a_+)$ and $(r,a_-)$ with $a_+ a_- =1$, which means
that the mapping between the coordinate patches is $a_+\mapsto \frac{1}{a_-}$.

In the coordinates (\ref{raa_coord}) the Dirac potential (\ref{Dir_pol}) on
the respective coordinate patch becomes
\begin{align} 
   A_\pm = \mp \frac{iN}{2}\frac{1}{1+a_\pm\ab_\pm}\paraa{\ab_\pm da_\pm - a_\pm d\ab_\pm}
\end{align}
The local gauge transformation mapping $A_+(r,a_+)$ to $A_-(r,a_+)$ is 
$g_{+-}=\frac{1}{\overline{a}_+^N}$.
Thus, a holomorphic line bundle with curvature $F$ 
is composed of polynomials in $a_\pm$ of at least degree $N$.

From now on, we will only consider the first coordinate patch with $a=a_+$.
The quantized constraints (\ref{q_constr}) become
\begin{align} 
   \hat\kappa_a = \hbar \partial_{a} + \frac{N}{2}\frac{\ab}{1+a\ab}, \qquad
   \hat\kappa_{\ab} = \hbar \partial_{\ab} + \frac{N}{2}\frac{a}{1+a\ab}, \qquad
   \hat\kappa_r =  \partial_{r} 
\end{align}
and the first two constraints can be expressed with the Kähler potential
\begin{align} 
   \phi = \ln \paraa{1+a\ab}^{\frac{N}{2}}
\end{align}
For the quantizations scheme (\ref{sip}), the constraint $\hat\kappa_r$
automatically becomes $0$. Due to the gauge invariance with respect to $r$,
the spheres of different radius are identified with each other.

The constants $c_n$ (\ref{cn_def}) become
\begin{align} 
  c_n = \int_0^\infty dr \, r^{2n+1} \frac{1}{(1+r^2)^\frac{N}{\hbar}} = \frac{1}{2}n!\frac{\Gamma(\frac{N}{\hbar}-n-1)}{\Gamma(\frac{N}{\hbar})}
\end{align}
This can be shown by substituting $u=r^2$ on the left hand side,
which results in the Beta function that evaluates to the right hand side. When 
$\frac{N}{\hbar}$ is integer, the series terminates when $n > \frac{N}{\hbar}-1$, since 
then the Gamma function has singularities. Otherwise, the constants $c_n$ exist
for all $n>0$, however we will exclude these infinite dimensional representations in the following.

From (\ref{raising_op}, \ref{lowering_op}) follows that
\begin{align} 
   \hat{a}\varphi_n = \sqrt{\frac{n+1}{\frac{N}{\hbar}-(n+1)}} \varphi_{n+1} , \qquad
   \hat{a}^\dagger \varphi_n = \sqrt{\frac{n}{\frac{N}{\hbar}-n}} \varphi_{n-1}
\end{align}
When we identify $\frac{N}{\hbar}= 2J+1$ and $m=J-n$ and we
substitute $\ket{m}=\varphi_n$, then the operators
\begin{align} 
   J_+  = \frac{N}{\hbar} \hat{a}^\dagger (1+ \hat{a}\hat{a}^\dagger)^{-1}, \qquad
   J_-  = \frac{N}{\hbar}  (1+ \hat{a}\hat{a}^\dagger)^{-1} \hat{a}
\end{align}
(compare (\ref{back_trans})) fulfill
\begin{align} 
   J_+  \ket{m}  =\sqrt{J(J+1) - m(m+1)} \ket{m} , \qquad
   J_-  \ket{m}  =\sqrt{J(J+1) - m(m -1 )} \ket{m} 
\end{align}
i.e. are the spin $J$ raising and lowering operators. This confirms that we indeed have
obtained a fuzzy sphere as the quantized phase space. 

\section{Comparision to Connes' approach}

In \cite{Con94} a possibly noncommutative $C^*$-algebra $C^*(V,F)$ is defined
for a foliation $F$ on a compact manifold $V$.

$C^*(V,F)$ is the completion of an algebra of functions on the holonomy groupoid
of the foliation. The holonomy gropoid is based on paths, which interconnect points
in the foliation and which are in parallel to the foliation. This means that there are solely
paths, which interconnect points, which are both within the same leave. In the holonomy gropoid
paths are identified with an equivalence relation, which have the same holonomy with
respect to parallel transport. For functions on the holonomy groupoid a convolutional product is
defined, for which a transverse measure of the foliation is used.
It is shown that when the foliation $F$ is based on a submersion $p:V\rightarrow B$, so that the leaves are 
$p^{-1}(x), x\in B$, the algebra $C^*(V,F)$ is isomorphic to $C_0(B)$, the algebra of 
continuous functions vanishing at $\infty$ on $B$.

In the approach presented herein, all points of a leave of the foliation are identified by gauge invariance, which
means that there is a path within one leave connecting the points. The noncommutative
product is defined on an algebra of functions on the space of leaves. This is analogous to
\cite{Con94}. However, we presented several examples based on a submersion
of a plane ($\reals^3\rightarrow \reals^2$) or a sphere $\reals^3 \setminus \{0\}\rightarrow S_2$,
which result in a noncommutative algebra, demonstrating that the present approach is different from
the one in \cite{Con94}.

\section{Discussion}

In this work, we have established a connection between constrained Hamiltonian systems
and noncommutative geometry by analyzing the strong-field limit of a charged particle
in an electromagnetic field. 

A distinctive feature of the system is that the classification of constraints
as first-class or second-class varies from point to point, depending on the local rank
of the field strength $F_{ij}$. In regions where $F$ has maximal rank, all constraints
are second-class, and the physical degrees of freedom are fully determined. In regions where
the rank of $F$ drops, some constraints become first-class, generating local gauge symmetries
that identify points along the leaves of the null foliation. This situation differs from standard gauge theories,
where the gauge group acts uniformly throughout spacetime. Here, the gauge symmetry is encoded
in a Lie algebroid rather than a Lie algebra, with the structure varying according to the geometry
of the field strength. 

The system (\ref{lim_action}) serves as a tractable toy model for general covariant theories.
Like general relativity, it possesses a vanishing Hamiltonian, with dynamics governed
entirely by constraints. The absence of a preferred time parameter reflects the
reparametrization invariance of the action. However, the present
system is simple enough to permit explicit quantization. The physical state space can be constructed
directly by imposing the quantum constraints, and the resulting Hilbert space has a clear geometric
interpretation as the quantization of the space of leaves of the null foliation. 

Perhaps the most striking result in this work is the natural emergence of noncommutative geometry
from the quantization of a constrained Hamiltonian system. The Dirac bracket (\ref{DB}) implies that
the coordinate functions on the physical state space do not Poisson-commute, with their bracket
given by the inverse of the field strength. Upon quantization, this translates into noncommuting
coordinate operators.

The examples illustrate how different field configurations lead to different noncommutative geometries.
In particular, the magnetic monopole field strength produces a fuzzy sphere, demonstrating that compact
noncommutative spaces arise naturally from topologically nontrivial field configurations.

In conclusion, the strong-field limit of a charged particle in an electromagnetic field provides a rich and tractable model that
bridges constrained Hamiltonian dynamics, singular foliations, and noncommutative geometry.

\end{document}